# Nonvolatile Resistive Switching in Nanocrystalline Molybdenum Disulfide with Ion-Based Plasticity

*Melkamu Belete[1], Satender Kataria[1,\*], Aykut Turfanda[1], Sam Vaziri[2], Thorsten Wahlbrink[3], Olof Engström[3] , Max C. Lemme[1,3,\*]*

[1]*RWTH Aachen University, Chair of Electronic Devices (ELD), Otto-Blumenthal-Str. 2, 52074 Aachen, Germany*

[2]*Stanford University, Department of Electrical Engineering, Stanford, CA 94305, USA*

[3]*AMO GmbH, Advanced Microelectronic Center Aachen, Otto-Blumenthal-Str. 25, 52074 Aachen, Germany*

\* satender.kataria@eld.rwth-aachen.de; max.lemme@eld.rwth-aachen.de

**Abstract:**

Non-volatile resistive switching is demonstrated in memristors with nanocrystalline molybdenum disulfide ($MoS_2$) as the active material. The vertical heterostructures consist of silicon, vertically aligned $MoS_2$ and chrome / gold metal electrodes. Electrical characterizations reveal a bipolar and forming free switching process with stable retention for at least 2500 seconds. Controlled experiments carried out in ambient and vacuum conditions suggest that the observed resistive switching is based on hydroxyl ions ($OH^-$). These originate from catalytic splitting of adsorbed water molecules by $MoS_2$. Experimental results in combination with analytical simulations further suggest that electric field driven movement of the mobile $OH^-$ ions along the vertical $MoS_2$ layers influences the energy barrier at the Si/$MoS_2$ interface. The scalable and semiconductor production compatible device fabrication process used in this work offers the opportunity to integrate such memristors into existing Si technology for future neuromorphic applications. The observed ion-based plasticity may be exploited in ionic-electronic devices based on TMDs and other 2D materials for memristive applications.

# 1. Introduction

Next generation computing is strongly believed to be the way forward to accommodate the vastly growing demands on computing, storage and communication.[1] Neuromorphic computing falls into this category and envisions the realization of artificial neural systems using electronic circuits to mimic the functioning of the human brain.[2–4] A neuromorphic computer, such is the hope, should be able to emulate the low power operation, highly efficient processing and real-time complex multi-tasking (parallel information processing) capabilities of the brain.[5,6] Significant research efforts are directed towards imitating neural synaptic plasticity, which is a feature that enables the brain to learn and process new information.[7–11] Electronic devices that can mimic this feature even to a certain extent are becoming highly desirable. Memristors (memory resistors), which are two-terminal passive circuit elements, are seen as primary candidates owing to the reversible resistive switching (RS) behavior they exhibit.[12,13] Memristors typically consist of metal-insulator-metal (MIM) structures, in which the insulator acts as the RS medium. The RS media of conventional MIM memristors are often based on phase change materials[14,15] or transition-metal oxides (TMOs).[13,14,16–18] The latter form and retract conductive filaments (CF) to trigger the RS effect and often require high currents initially to form the filaments. Recently, two-dimensional (2D) transition metal dichalcogenides (TMDs) have also been investigated for their potential use as a RS medium in memristor devices.[19–30] So far, mostly planar or lateral devices based on horizontally aligned TMD layers obtained either by mechanical exfoliation or chemical vapor deposition have been studied. Molybdenum disulfide ($MoS_2$) is a prominent member of the 2D TMD family that is getting considerable attention on account of its interesting electronic properties including a thickness-dependent band gap [31–33] and memristive behavior. The latter has so far been attributed to bias-induced migration of sulfur vacancies and grain boundaries in the case of single layer $MoS_2$[19,22], and lattice distortion, reversible modulation of $MoS_2$ phases and migration of oxygen ions in multilayers.[20,27,29] Nevertheless, the origin and mechanism of the RS behavior of MoS$_2$ is still under debate.

Recently, we demonstrated the presence of OH$^-$ and their field-induced movement in vapor-phase grown MoS$_2$ with vertically aligned layers.[34,35] Here, we demonstrate the existence and investigate the origin of nonvolatile resistive switching in silicon (Si) / MoS$_2$ / metal heterostructure devices, where MoS$_2$ with vertically aligned layers serves as the RS medium. In analogy to the conventional MIM configuration, we name the devices in this work semiconductor / semiconductor / metal (SSM) memristors. The OH$^-$ ion-mediated RS behavior is observed in vertically aligned MoS$_2$ layers in a vertical device architecture, which is preferred for practical applications due to its potential for ultimately small footprints.[23] The fabrication process used in this work is scalable and semiconductor manufacturing compatible and includes the growth of large-area MoS$_2$ directly on Si. This is an important aspect towards future integration of such devices with the existing Si-technology platform.

## 2. Results and Discussion

Direct current (DC) current-voltage (I-V) and current-time (I-t) measurements were conducted on the SSM memristors in a Lakeshore cryogenic probe station using a Keithley 4200SCS parameter analyzer. The DC measurements were conducted in both ambient and vacuum conditions, at room temperature. Voltage was applied to the top electrode ($V_{TE}$) while keeping the Si grounded as illustrated in the wiring setup shown in the inset of **Figure 1a**. First, different DC sweep ranges were applied to determine the window of operation for RS. It was found that the switching effect in the negative bias regime becomes observable only for $V_{TE} \leq -4$ V and generally scales with the magnitude of the applied bias (supporting information Figure S1). To avert potential device damage caused by voltage stress during measurements, further I-V measurements were limited to $\pm 5$ V. Figure 1a shows a representative I-V characteristics of the SSM memristors in ambient conditons. It exhibits a forming free and bipolar RS behavior, in which the resistance state is modulated between two resistance levels, namely the high resistance state (HRS) and the low resistance state (LRS). Transition from the HRS to LRS (SET process) is observed in the negative bias regime, while the reverse transition from the LRS to HRS (RESET process) happens in the positive bias regime. This bipolar switching process is similar to valence change switching in transition metal oxides, albeit different underlying mechanisms apply.[36] The I-V measurements were carried out in the following sequence. First, $V_{TE}$ was swept from 0 to -5 V (sweep 1), during which the onset of the SET process is observed at about -4 V. As a result, the device switches from the HRS to the LRS leading to the observed current increase. Next, $V_{TE}$ was swept from -5 V back to 0 V (sweep 2) during which the device remains in the LRS. Afterwards, $V_{TE}$ was swept from 0 V to +5 V (sweep 3) and finally back to 0 V (sweep 4). The device maintains the LRS during sweep 3, but it switches to the HRS during sweep 4. It is observed that the switching ratio of the RESET process is much smaller than that of the SET process. In addition to the observed hysteretic behavior, which is the main focus of this work, the I-V characteristics in Figure 1a also exhibit asymmetric behavior indicating that the charge carrier transport in the present devices is mainly determined by the interface barriers. Zhu et al. have reported similar I-V characteristics. i.e. a larger switching ratio in the SET than in the RESET regime for Au/$Li_xMoS_2$/Au lateral structures and have demonstrated RS phenomenon by field-induced migration of lithium ions ($Li^+$) in $MoS_2$ layers.[27]

In the present devices, the observed RS effect can be explained by movement of $OH^-$ ions inside the layered $MoS_2$. Such ions have been experimentally observed in a previous study through analytical methods and eletrical measurements.[34] There, it was also demonstrated that the $OH^-$ ions move presumably along the van der Waals (vdW) gaps towards (away from) the Si/MoS$_2$ interface in response to a negative (positive) bias applied on the top electrode.[34,35] These ions are expected to originate from water molecules that have been split by the catalytic effect of MoS$_2$ on water.[37–39] This is in line with other reports where the layered crystal structure of MoS$_2$ and the associated anisotripic electronic properties facilitate ion transport along the layers.[27,40,41]

To visualize how the RS effect takes place in the present system, schematic band diagrams are provided in Figure 1b-e. The figure illustrates alignments of the bands during thermal equilibrium (Figure 1b), the SET process (Figure 1c), the RESET process (Figure 1d) and the READ (Figure 1e) conditions. For the present devices, this means that a negative $V_{TE}$ (SET) pushes the OH$^-$ ions towards the Si/MoS$_2$ interface and lowers the energy barrier at the interface compared to a situation where the ions are located farther away from the Si/MoS$_2$ interface. As a result, the devices switch to the LRS (red curve in Figure 2). On the other hand, the OH$^-$ ions move away from the Si/MoS$_2$ interface when a positive $V_{TE}$ is applied (RESET). Now, the MoS$_2$ band aligment results in a relatively higher interface barrier. Hence, the positive bias switches the device resistance to the HRS (black curve in Figure 2). During READ, the bands align in such a way that holes move from Si to MoS$_2$ overcoming the hole barrier at the Si/MoS$_2$ interface, while electrons move in the opposite direction. Due to the high barrier encountered by electrons at the Cr/MoS$_2$ interface in this particular configuration (Figure 1e), the RS in the devices is dominated by hole transport, which in turn is controlled by the Si/MoS$_2$ hole barrier. The field-driven movement of the OH$^-$ ions thus tunes the energy barrier at the Si/MoS$_2$ interface and therefore controls the charge carrier transfer from Si to MoS$_2$. This in turn gives rise to modulation of the device's resistance between LRS and HRS and hence the observed memristive behavior.

Non-volatility of RS, and in particular endurance and state-retention, are important figures of merit for device applicability. Endurance is a measure of how many times a memristor device can be switched between the LRS and HRS while maintaining a reasonable switching ratio.[13,21] State-retention is also commonly used to test if the resistance states are stable over a period of time, following the SET- and RESET transitions.[13,21] To achieve reliable results during endurance tests, determining the right programming time and programming voltages for the SET and RESET processes is very crucial and can only be done empirically. Initial experiments to this end can be found in the supporting information (Figure S2). After obtaining appropriate measurement parameters, we carried out endurance tests for 140 manual DC switching cycles in ambient conditions. The endurance data was acquired by reading out the resistance values from the corresponding current levels at -1.5 V during each switching cycle (READ). Each READ was done immediately after a SET and RESET at programming voltages of -3.5 V and +4 V, respectively, which were applied as step voltage signals held for 2 s. The results obtained are shown in Figure 2a, where a clear RS behavior is evident for 140 manual DC switching cycles. The endurance data exhibits a distinct resistance drift in the beginning. This may be attributed to an initial reduction of negative ions inside $MoS_2$ due to their interaction with holes in the Si or to an intial redistribution of the ions from an equal distribution to a more concentrated (Gaussian) distribution due to the electric fields. The data also shows a less pronounced resistance drift in further cycles, which requires future investigation. As shown in the supporting information (Figure S3a and S3c), similar drift characeistics were also observed in endurance measurements on other devices. Despite the general similarity, slight variations in the absolute resistance values among different devices are noticed. Such a cell-to-cell (spatial) variability is in fact a major challenge for RS devices in general, and it is believed to hinder their use for computing and memory applications.[13] In 2D TMD memristors, this may be partially attributed to the lack of spatial homogeneity in the large area TMD films[22], an issue that can be expected to be solved by further maturing the manufacturing technology.

State retention measurements were also performed in ambient atmosphere according to the following procedure. First, a SET voltage of -4 V was applied for 2 s to switch the device to the LRS, followed by 400 subsequent READ operations at -1.5 V. Next, immediately after the 400th readout cycle in the LRS, a RESET voltage of +4 V was applied for 2 s to switch to the HRS. Then, another 400 subsequent READ operations were performed in the HRS. Figure 2b shows the resulting retention data demonstrating that the devices are able to retain the resistance states for at least 2500 seconds, albeit with a slight resistance shift in the LRS after 1900 seconds (after the 300th readout cycle). This shift appears to be random and is not fully understood. It may be due to a parasitic influence on the OH- ions during the READ phase, which can be addressed with a more suitable READ voltage in the future. Nevertheless, the overall stable state retention data suggests that the mobile OH- ions dictating the RS effect drift substantially only during the programming phases, and not during the READ phase.

We peformed analytical simulations that take into account the position of the ions within the $MoS_2$ and found that it determines the electric field distribution in the structures. This in turn influences the $MoS_2$ band alignment with respect to Si and therefore the energy barriers at the Si/$MoS_2$ interface. Figure 2c illustrates an assumed Gaussian distribution of OH- ions in $MoS_2$ at a randomly selected positions for three cases: (i) starting position with no bias (blue), (ii) a position shifted towards the Si/$MoS_2$ interface for negative bias (red) and (iii) a position shifted away from the interface for positive bias (black). Calculations of the corresponding $MoS_2$ valence bands ($E_V$) demonstrate that the energy barrier at the Si/$MoS_2$ interface is modulated by the position of the OH- ions with respect to the interface (Figure 2d). This influences the charge carrier transfer from Si to $MoS_2$ and gives rise to different resistance states depending on the position of the mobile charges. The fundamentals and details behind the analytical calculations can be found in.[42]

We also investigated the influence of voltage sweep rates on the RS behavior and noticed an increase in the switching ratio and a decrease in the onset-voltage of the SET process for a decreasing sweep rate (Figure 3a). A similar observation was reported by Ge et. al, who attributed the decrease in SET voltage to an enhanced ionic diffusion as a result of the additional time that the slower sweep rate offers. [23]

Measurements in vacuum at $7 \times 10^{-5}$ mbar were carried out to confirm the hypothesis that the driving force behind the RS phenomenon originates from catalytically split water adsorbates. The devices were kept in the vacuum chamber with active pumping for about 64 hours prior to the measurements to ensure that water molecules were extracted efficiently. Figure 3b shows I-V characteristics measured in vacuum under the same biasing conditions as used in ambient. A device in vacuum exhibits very low switching ratios compared to ambient conditions (see Figure 1a). This is confirmed by endurance data in vacuum, which also shows a negligible difference between the HRS and LRS (Figure 3c). Similar trends were obtained from endurance tests carried out in vacuum conditions on other devices, as shown in the supporting information (Figure S3b and S3d).

Finally, we conducted an experiment where endurance tests were carried out first in ambient, then in vacuum and then again in ambient conditions on the same device. As results in Figure. 3d - f show that, the strong RS effect is restored in ambient condition after it disappeared in vacuum, affirming our initial premise. There exist very few experimental reports on RS of two-terminal $MoS_2$ memristors with a vertical architecture in vacuum conditions. Sangwan et al. have reported gate-modulated RS in vacuum of single layer polycrystalline

MoS$_2$-based three-terminal memtransistors with a lateral arctechture and attributed the RS effect to migration of grain boundaries and sulfur vacancies.[19,22] The absence of considerable switching windows in our measurements under vacuum, however, rules out a dominating role of grain boundaries and sulfur vacancies in the RS. Kalita et al. have reported a volatile threshold switching effect in graphene / vertical MoS$_2$ / Ni structures that is reduced in vacuum. They attributed the effect to oxygen in ambient conditions [29], different to the results and conclusions presented here.

## 3. Conclusion

Vertical Si / MoS$_2$ / Cr memristors with vertically aligned MoS$_2$ layers grown directly on Si are fabricated using scalable and semiconductor production compatible processes. Non-volatile resistive switching behavior is demonstrated in these devices and the main origin of the phenomenon is investigated. DC I-V characterization results showed forming-free bipolar RS, in which the SET and RESET transitions are achieved at negative and positive biases, respectively. Endurance and state-retention tests performed on the devices demonstrate that the observed non-volatile RS behavior is quite stable over a period of time despite successive biasing. Comparison of measurements carried out in ambient and vacuum conditions provide plausible arguments that the observed RS effect is due to OH$^-$ ions that stem from catalytic splitting of adsorbed water molecules, in agreement with previous findings.[34] Analytical simulations support experiments and suggest that the movement of such ions towards (away from) the Si/MoS$_2$ interface in response to negative (positive) electric-fields dictates the observed RS behavior by tuning the energy barriers at the interface. The vertically aligned MoS$_2$ layers on Si are favorable for vertical crossbar device architectures for memristive applications. Furthermore, the scalable fabrication processes employed in this work lays out a path for easy integration of novel 2D materials-based memristors with existing semiconductor technology.

## 4. Experimental Section

To fabricate the SSM heterostructure memristors, $MoS_2$ films were grown by thermally assisted conversion of molybdenum under sulfur atmosphere directly on p-type Si (100) substrates. Details of the $MoS_2$ synthesis process are described in the supporting information and in.[34] After $MoS_2$ growth, top electrodes were defined on the $MoS_2$ films though a photolithography step followed by evaporation of 20 nm chromium (Cr) and 120 nm gold (Au) and a lift-off process. Next, a Cr/Au metal back contact was formed on the Si substrates through evaporation. In the resultant SSM memristors, vertically aligned $MoS_2$ is sandwiched between Cr and Si. The complete fabrication process scheme and a top-view scanning electron microscope image are provided in the supporting information (Figure S4a and S4b). The $MoS_2$ phase formation in the as-grown films was confirmed using micro Raman spectroscopy at a laser wavelength of 532 nm. Figure 4a shows the acquired Raman spectrum with the two prominent peaks corresponding to the $E^1_{2g}$ and $A_{1g}$ modes of 2H-$MoS_2$ phase.[43] The peak intensity of the $A_{1g}$ mode is nearly four times higher than that of $E^1_{2g}$, indicating the formation of vertically aligned $MoS_2$ layers.[44,45] Atomic force microscopy (AFM) was used to measure the thickness of the as-grown films, which is found to be ~15 nm (Figure 4b). Transmission electron microscopy (TEM) was employed to investigate the cross-section of the device structure. The cross-sectional TEM image in Figure 4c shows predominantly vertical orientation of $MoS_2$ layers with respect to the Si substrate. This is in agreement with the Raman studies where the vertical orientation of the $MoS_2$ layers is manifested by the large $A_{1g}/E^1_{2g}$ peak intensity ratio (Figure 4a). The TEM images also reveal a ~ 2.5 nm interfacial silicon oxide layer (IL) between Si and $MoS_2$. This IL is presumably very leaky compared to standard silicon dioxide ($SiO_2$) of similar thickness at comparable bias level[46] and is therefore considered nearly transparent to electrons under DC bias. Hence, the charge carrier transport between Si and MoS$_2$ is nearly unaffected by the presence of this layer. A schematic of the device structure is depicted in Figure 4d.

**Supporting Information**

Supporting Information is available from the Wiley Online Library or from the author.

**Acknowledgements**

We would like to thank Gregor Schulte (University of Siegen) for his help in depositing the initial molybdenum films and Prof. Joachim Mayer and Maximilian Kruth for TEM imaging. We also thank Dr. Daniel Neumeier (AMO GmbH) for fruitful discussions. Financial support from the European Commission is gratefully acknowledged (Graphene Flagship, 785219, QUEFORMAL, 829035).

**References**

(1) Murugesan, S. and Colwell, B. Next-Generation Computing Paradigms. *IEEE Computer Society* 2016, *49* (09), 14–20. https://doi.org/10.1109/MC.2016.265.

(2) Mead, C. Neuromorphic Electronic Systems. *Proceedings of the IEEE* 1990, *78* (10), 1629–1636. https://doi.org/10.1109/5.58356.

(3) Monroe, D. Neuromorphic Computing Gets Ready For the (Really) Big Time https://cacm.acm.org/magazines/2014/6/175183-neuromorphic-computing-gets-ready-for-the-really-big-time/fulltext (accessed May 7, 2019).

(4) Zhao, W. S.; Agnus, G.; Derycke, V.; Filoramo, A.; Bourgoin, J.-P.; Gamrat, C. Nanotube Devices Based Crossbar Architecture: Toward Neuromorphic Computing. *Nanotechnology* 2010, *21* (17), 175202. https://doi.org/10.1088/0957-4484/21/17/175202.

(5) Laughlin, S. B.; Steveninck, R. R. de R. van; Anderson, J. C. The Metabolic Cost of Neural Information. *Nature Neuroscience* 1998, *1* (1), 36–41. https://doi.org/10.1038/236.

(6) Tian, H.; Zhao, L.; Wang, X.; Yeh, Y.-W.; Yao, N.; Rand, B. P.; Ren, T.-L. Extremely Low Operating Current Resistive Memory Based on Exfoliated 2D Perovskite Single Crystals for Neuromorphic Computing. *ACS Nano* 2017, *11* (12), 12247–12256. https://doi.org/10.1021/acsnano.7b05726.

(7) Martin, S. J.; Grimwood, P. D.; Morris, R. G. M. Synaptic Plasticity and Memory: An Evaluation of the Hypothesis. *Annual Review of Neuroscience* 2000, *23* (1), 649–711. https://doi.org/10.1146/annurev.neuro.23.1.649.

(8) Zucker, R. S.; Regehr, W. G. Short-Term Synaptic Plasticity. *Annual Review of Physiology* 2002, *64* (1), 355–405. https://doi.org/10.1146/annurev.physiol.64.092501.114547.

(9) McAllister, A. K.; Katz, L. C.; Lo, D. C. Neurotrophins and Synaptic Plasticity. *Annual Review of Neuroscience* 1999, *22* (1), 295–318. https://doi.org/10.1146/annurev.neuro.22.1.295.

(10) Holtmaat, A.; Svoboda, K. Experience-Dependent Structural Synaptic Plasticity in the Mammalian Brain. *Nature Reviews Neuroscience* 2009, *10* (9), 647–658. https://doi.org/10.1038/nrn2699.

(11) Ge, S.; Yang, C.; Hsu, K.; Ming, G.; Song, H. A Critical Period for Enhanced Synaptic Plasticity in Newly Generated Neurons of the Adult Brain. *Neuron* 2007, *54* (4), 559–566. https://doi.org/10.1016/j.neuron.2007.05.002.

(12) Maan, A. K.; Jayadevi, D. A.; James, A. P. A Survey of Memristive Threshold Logic Circuits. *IEEE Transactions on Neural Networks and Learning Systems* 2017, *28* (8), 1734–1746. https://doi.org/10.1109/TNNLS.2016.2547842.

(13) Lanza, M.; Wong, H.-S. P.; Pop, E.; Ielmini, D.; Strukov, D.; Regan, B. C.; Larcher, L.; Villena, M. A.; Yang, J. J.; Goux, L.; et al. Recommended Methods to Study Resistive Switching Devices. *Advanced Electronic Materials* 2019, *5* (1), 1800143. https://doi.org/10.1002/aelm.201800143.

(14) Wouters, D. J.; Waser, R.; Wuttig, M. Phase-Change and Redox-Based Resistive Switching Memories. *Proceedings of the IEEE* 2015, *103* (8), 1274–1288. https://doi.org/10.1109/JPROC.2015.2433311.

(15) Wuttig, M.; Yamada, N. Phase-Change Materials for Rewriteable Data Storage. *Nature Materials* 2007, *6* (11), 824–832. https://doi.org/10.1038/nmat2009.

(16) Zhang, M.; Long, S.; Li, Y.; Liu, Q.; Lv, H.; Miranda, E.; Suñé, J.; Liu, M. Analysis on the Filament Structure Evolution in Reset Transition of Cu/HfO2/Pt RRAM Device. *Nanoscale Res Lett* 2016, *11 (269)*. https://doi.org/10.1186/s11671-016-1484-8.

(17) Wong, H.-P.; Lee, H.; Yu, S.; Chen, Y.; Wu, Y.; Chen, P.; Lee, B.; Chen, F. T.; Tsai, M. Metal–Oxide RRAM. *Proceedings of the IEEE* 2012, *100* (6), 1951–1970. https://doi.org/10.1109/JPROC.2012.2190369.

(18) Sawa, A. Resistive Switching in Transition Metal Oxides. *Materials Today* 2008, *11* (6), 28–36. https://doi.org/10.1016/S1369-7021(08)70119-6.

(19) Sangwan, V. K.; Jariwala, D.; Kim, I. S.; Chen, K.-S.; Marks, T. J.; Lauhon, L. J.; Hersam, M. C. Gate-Tunable Memristive Phenomena Mediated by Grain Boundaries in Single-Layer $MoS_2$. *Nature Nanotechnology* 2015, *10* (5), 403–406. https://doi.org/10.1038/nnano.2015.56.

(20) Cheng, P.; Sun, K.; Hu, Y. H. Memristive Behavior and Ideal Memristor of 1T Phase MoS2 Nanosheets. *Nano Lett.* 2016, *16* (1), 572–576. https://doi.org/10.1021/acs.nanolett.5b04260.

(21) Hui, F.; Grustan-Gutierrez, E.; Long, S.; Liu, Q.; Ott, A. K.; Ferrari, A. C.; Lanza, M. Graphene and Related Materials for Resistive Random Access Memories. *Advanced Electronic Materials* 2017, *3* (8), 1600195. https://doi.org/10.1002/aelm.201600195.

(22) Sangwan, V. K.; Lee, H.-S.; Bergeron, H.; Balla, I.; Beck, M. E.; Chen, K.-S.; Hersam, M. C. Multi-Terminal Memtransistors from Polycrystalline Monolayer Molybdenum Disulfide. *Nature* 2018, *554* (7693), 500–504. https://doi.org/10.1038/nature25747.

(23) Ge, R.; Wu, X.; Kim, M.; Shi, J.; Sonde, S.; Tao, L.; Zhang, Y.; Lee, J. C.; Akinwande, D. Atomristor: Nonvolatile Resistance Switching in Atomic Sheets of Transition Metal Dichalcogenides. *Nano Lett.* 2018, *18* (1), 434–441. https://doi.org/10.1021/acs.nanolett.7b04342.

(24) Kim, M.; Ge, R.; Wu, X.; Lan, X.; Tice, J.; Lee, J. C.; Akinwande, D. Zero-Static Power Radio-Frequency Switches Based on MoS 2 Atomristors. *Nature Communications* 2018, *9* (1), 2524. https://doi.org/10.1038/s41467-018-04934-x.

(25) Li, D.; Wu, B.; Zhu, X.; Wang, J.; Ryu, B.; Lu, W. D.; Lu, W.; Liang, X. MoS2 Memristors Exhibiting Variable Switching Characteristics toward Biorealistic Synaptic Emulation. *ACS Nano* 2018, *12* (9), 9240–9252. https://doi.org/10.1021/acsnano.8b03977.

(26) Wang, M.; Cai, S.; Pan, C.; Wang, C.; Lian, X.; Zhuo, Y.; Xu, K.; Cao, T.; Pan, X.; Wang, B.; et al. Robust Memristors Based on Layered Two-Dimensional Materials. *Nat Electron* 2018, *1* (2), 130–136. https://doi.org/10.1038/s41928-018-0021-4.

(27) Zhu, X.; Li, D.; Liang, X.; Lu, W. D. Ionic Modulation and Ionic Coupling Effects in MoS 2 Devices for Neuromorphic Computing. *Nature Materials* 2019, *18* (2), 141. https://doi.org/10.1038/s41563-018-0248-5.

(28) Zhang, F.; Zhang, H.; Krylyuk, S.; Milligan, C. A.; Zhu, Y.; Zemlyanov, D. Y.; Bendersky, L. A.; Burton, B. P.; Davydov, A. V.; Appenzeller, J. Electric-Field Induced Structural Transition in Vertical MoTe 2 - and Mo 1–x W x Te 2 -Based Resistive Memories. *Nature Materials* 2019, *18* (1), 55–61. https://doi.org/10.1038/s41563-018-0234-y.

(29) Kalita, H.; Krishnaprasad, A.; Choudhary, N.; Das, S.; Dev, D.; Ding, Y.; Tetard, L.; Chung, H.-S.; Jung, Y.; Roy, T. Artificial Neuron Using Vertical MoS 2 /Graphene Threshold Switching Memristors. *Scientific Reports* 2019, *9* (1), 53. https://doi.org/10.1038/s41598-018-35828-z.

(30) Kumar, M.; Ban, D.-K.; Kim, S. M.; Kim, J.; Wong, C.-P. Vertically Aligned WS2 Layers for High-Performing Memristors and Artificial Synapses. *Advanced Electronic Materials* 2019, *0* (0), 1900467. https://doi.org/10.1002/aelm.201900467.

(31) Yazyev, O. V.; Kis, A. MoS2 and Semiconductors in the Flatland. *Materials Today* 2015, *18* (1), 20–30. https://doi.org/10.1016/j.mattod.2014.07.005.

(32) Mak, K. F.; Lee, C.; Hone, J.; Shan, J.; Heinz, T. F. Atomically Thin MoS2: A New Direct-Gap Semiconductor. *Phys. Rev. Lett.* 2010, *105* (13), 136805. https://doi.org/10.1103/PhysRevLett.105.136805.

(33) Chhowalla, M.; Shin, H. S.; Eda, G.; Li, L.-J.; Loh, K. P.; Zhang, H. The Chemistry of Two-Dimensional Layered Transition Metal Dichalcogenide Nanosheets. *Nature Chemistry* 2013, *5* (4), 263–275. https://doi.org/10.1038/nchem.1589.

(34) Belete, M.; Kataria, S.; Koch, U.; Kruth, M.; Engelhard, C.; Mayer, J.; Engström, O.; Lemme, M. C. Dielectric Properties and Ion Transport in Layered MoS2 Grown by Vapor-Phase Sulfurization for Potential Applications in Nanoelectronics. *ACS Appl. Nano Mater.* 2018, *1* (11), 6197–6204. https://doi.org/10.1021/acsanm.8b01412.

(35) Belete, M.; Kataria, S.; Engström, O.; Lemme, M. C. Defects in Layered Vapor-Phase Grown MOS2. In *2017 75th Annual Device Research Conference (DRC)*; 2017; pp 1–2. https://doi.org/10.1109/DRC.2017.7999486.

(36) Jeong, D. S.; Schroeder, H.; Waser, R. Mechanism for Bipolar Switching in a Pt/TiO2/Pt Resistive Switching Cell. *Phys. Rev. B* 2009, *79* (19), 195317. https://doi.org/10.1103/PhysRevB.79.195317.

(37) Li, S.; Wang, S.; Salamone, M. M.; Robertson, A. W.; Nayak, S.; Kim, H.; Tsang, S. C. E.; Pasta, M.; Warner, J. H. Edge-Enriched 2D MoS2 Thin Films Grown by Chemical Vapor Deposition for Enhanced Catalytic Performance. *ACS Catal.* 2017, *7* (1), 877–886. https://doi.org/10.1021/acscatal.6b02663.

(38) Tributsch, H.; Bennett, J. C. Electrochemistry and Photochemistry of MoS2 Layer Crystals. I. *Journal of Electroanalytical Chemistry and Interfacial Electrochemistry* 1977, *81* (1), 97–111. https://doi.org/10.1016/S0022-0728(77)80363-X.

(39) Karunadasa, H. I.; Chang, C. J.; Long, J. R. A Molecular Molybdenum-Oxo Catalyst for Generating Hydrogen from Water. *Nature* 2010, *464* (7293), 1329–1333. https://doi.org/10.1038/nature08969.

(40) Stephenson, T.; Li, Z.; Olsen, B.; Mitlin, D. Lithium Ion Battery Applications of Molybdenum Disulfide (MoS2) Nanocomposites. *Energy Environ. Sci.* 2013, *7* (1), 209–231. https://doi.org/10.1039/C3EE42591F.

(41) Hong, J.; Li, K.; Jin, C.; Zhang, X.; Zhang, Z.; Yuan, J. Layer-Dependent Anisotropic Electronic Structure of Freestanding Quasi-Two-Dimensional $MoS_2$. *Phys. Rev. B* 2016, *93* (7), 075440. https://doi.org/10.1103/PhysRevB.93.075440.

(42) Belete, M.; Engstrom, O.; Vaziri, S.; Kataria, S.; Lemme, M. C. Electron Transport across Vertical Silicon / Molybdenum Disulfide (MoS2) / Graphene Heterostructures. *submitted*.

(43) Deokar, G.; Rajput, N. S.; Vancsó, P.; Ravaux, F.; Jouiad, M.; Vignaud, D.; Cecchet, F.; Colomer, J.-F. Large Area Growth of Vertically Aligned Luminescent MoS2 Nanosheets. *Nanoscale* 2016, *9* (1), 277–287. https://doi.org/10.1039/C6NR07965B.

(44) Kong, D.; Wang, H.; Cha, J. J.; Pasta, M.; Koski, K. J.; Yao, J.; Cui, Y. Synthesis of MoS2 and MoSe2 Films with Vertically Aligned Layers. *Nano Lett.* 2013, *13* (3), 1341–1347. https://doi.org/10.1021/nl400258t.

(45) Wang, H.; Kong, D.; Johanes, P.; Cha, J. J.; Zheng, G.; Yan, K.; Liu, N.; Cui, Y. MoSe2 and WSe2 Nanofilms with Vertically Aligned Molecular Layers on Curved and Rough Surfaces. *Nano Lett.* 2013, *13* (7), 3426–3433. https://doi.org/10.1021/nl401944f.

(46) Rana, F.; Tiwari, S.; Buchanan, D. A. Self-consistent Modeling of Accumulation Layers and Tunneling Currents through Very Thin Oxides. *Appl. Phys. Lett.* 1996, *69* (8), 1104–1106. https://doi.org/10.1063/1.117072.

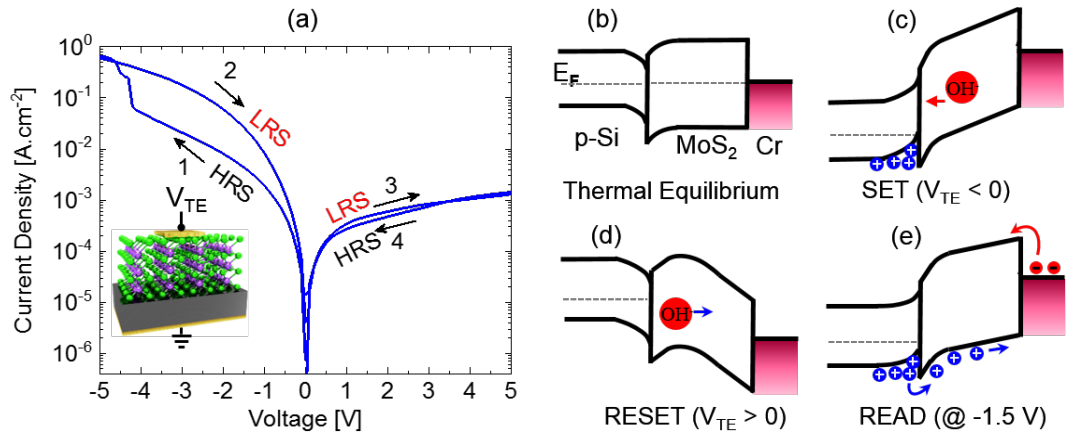

Figure 1- (a) Representative I-V characteristics of an SSM memristor measured in ambient condition showing bipolar resistive switching. The DC sweep was carried out as follows. Sweep 1: sweeping $V_{TE}$ from 0 to -5 V, during which the resistance shows a sudden increase at around -4 V indicating a HRS to LRS transition (SET). Sweep 2: $V_{TE}$ was swept from -5 V back to 0 and the device remained in LRS. Sweep 3: $V_{TE}$ was swept from 0 to +5 V and the device maintained the LRS. Sweep 4: $V_{TE}$ was swept from +5 V back to 0, during which a LRS to HRS transition is evident although the switching ratio is not as pronounced as in the SET. The inset shows the schematic of the device with the wiring setup. Schematic band diagrams of the current SSM device illustrating the aligment of the bands: (b) Band alignment at thermal equilibrium with negative charges (OH$^-$ ions) residing inside the MoS$_2$ bulk. (c) Band alignment during the SET process, during which negative bias pushes the OH$^-$ ions (red circle in the MoS$_2$ band gap) to the Si/MoS$_2$ interface as shown by the red arrow while the Si is in accumulation. (d) Band alignment after the RESET process by applying a positive bias. In this case, the OH$^-$ ions (red circle in the MoS$_2$ band gap) are pulled away from the Si/MoS$_2$ interface as indicated by the blue arrow while the Si is in depletion. (e) Band alignment during READ at -1.5 V.

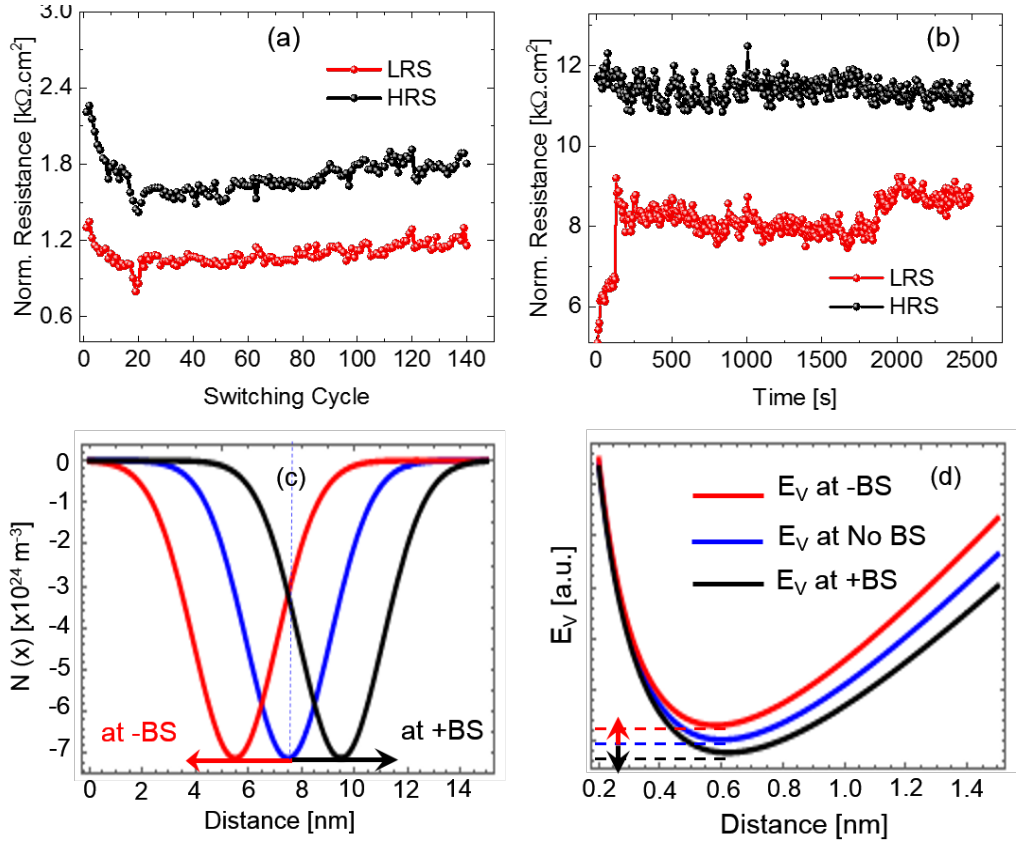

Figure 2- (a) Endurance data from an SSM memristor for 140 manual DC switching cycles performed in ambient conditions. The resistance values in the LRS and HRS of each cycle were determined from the current READ at -1.5 V immediately after applying the SET and RESET voltages of -3.5 V and +4 V, respectively, for 2 s. (b) State-retention measurement in ambient condition of another SSM memristor depicting stable retention of the RS states. A SET voltage of -4V was applied for 2 s on the tope electrode followed by 400 subsequent current readings at -1.5 V from which the LRS resistances were determined. The procedure was repeated after a RESET at +4V. The vertical axis shows the device's resistance normalized to the device area (200 x 200 μm$^2$) and the horizontal axis shows time, which is translated from the corresponding READ cycles accomplished in the LRS (HRS) after applying the negative (positive) programing voltages. Analytical calculations demonstrating the influence of electric field-induced movement of mobile OH$^-$ ions on the hole barrier at the Si/MoS$_2$ interface: (c) Gaussian distribution of the negative mobile charges inside the MoS$_2$ bulk at an initial position before biasing (blue), shifted to the Si/MoS$_2$ interface by negative bias (red) and shifted to the Cr/MoS$_2$ interface by positive bias (black). N(x) in the y-axis is the negative ion charge concentration and the x-axis is the distance between the Si/MoS$_2$ and Cr/MoS$_2$ interfaces, which basically is the same as the MoS$_2$ film thickness. (d) Analytical calculation showing the alignment of the MoS$_2$ valence band (E$_V$) for charge positions indicated in (c). In both graphs, the Si/MoS$_2$ interface is considered at the left vertical axis, while the Cr/MoS$_2$ interface is considered at the right vertical axis. The calculations show a decrease of the hole barrier at the Si/MoS$_2$ interface for negative bias as negative mobile charges are driven towards the interface and an increase during positive bias that pulls the mobile charges away from the interface. Hence, this influence on the barrier translates into the LRS for negative bias and HRS for positive bias.

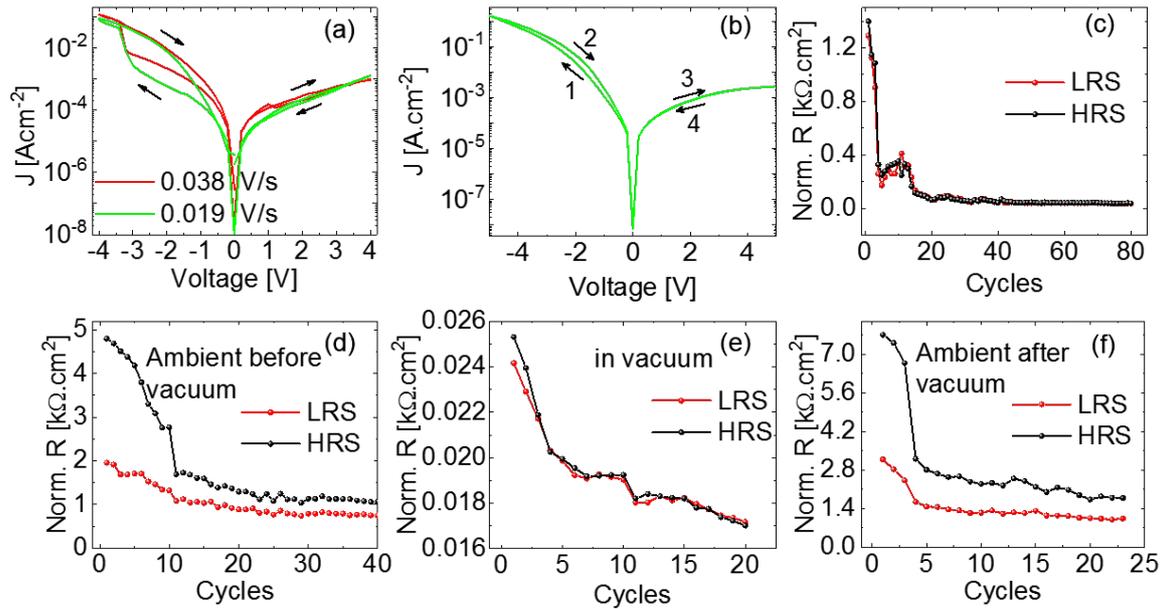

Figure 3- (a) I-V characteristics of an SSM memristor in ambient condition showing the effect of the sweep rate of the applied bias on the RS effect. The switching ratio of the SET increases for a decreasing sweep rate, while the decreasing rate reduces the onset-bias of the SET process. This supports the hypothesis of ionic transport inside $MoS_2$ playing a major role in the observed RS effect. (b) I-V characteristics of the SSM memristor measured in vacuum conditions ($7\times10^{-5}$ mbar). The data shows a dramatic decline in switching ratio compared to the corresponding result achieved in ambient conditions on the same device (Figure 2a), suggesting a considerable effect of adsorbents from the environment on the RS behavior. (c) Endurance test performed in vacuum conditions ($7\times10^{-5}$ mbar) on the same device tested in Figure 3a, showing a negligible RS behavior in the absence of adsorbents. The endurance data here is acquired following exactly the same procedure described in Figure 3a. Room temperature endurance measurements carried out on a single $MoS_2$-based SSM device under three conditions (i.e. ambient, vacuum and then again in ambient): (d) Endurance data for 100 manual DC switching cycles performed in ambient conditions showing a clear RS effect. (e) Endurance test of the same device in vacuum conditions showing negligible RS effect in the absence of adsorbents. The test was carried out immediately after the ambient measurements in "c". (f) Endurance test performed on the same device again in ambient condition after venting the chamber following the vacuum measurements in "d". The observation that the RS effect is much more pronounced in ambient conditions but disappears in vacuum supports a water adsorbent-driven process. The resistance values of the LRS and HRS in each cycle were determined by measuring the current as a function of time at a reading bias of -1.5 V immediately after applying the set/reset programming voltages, respectively.

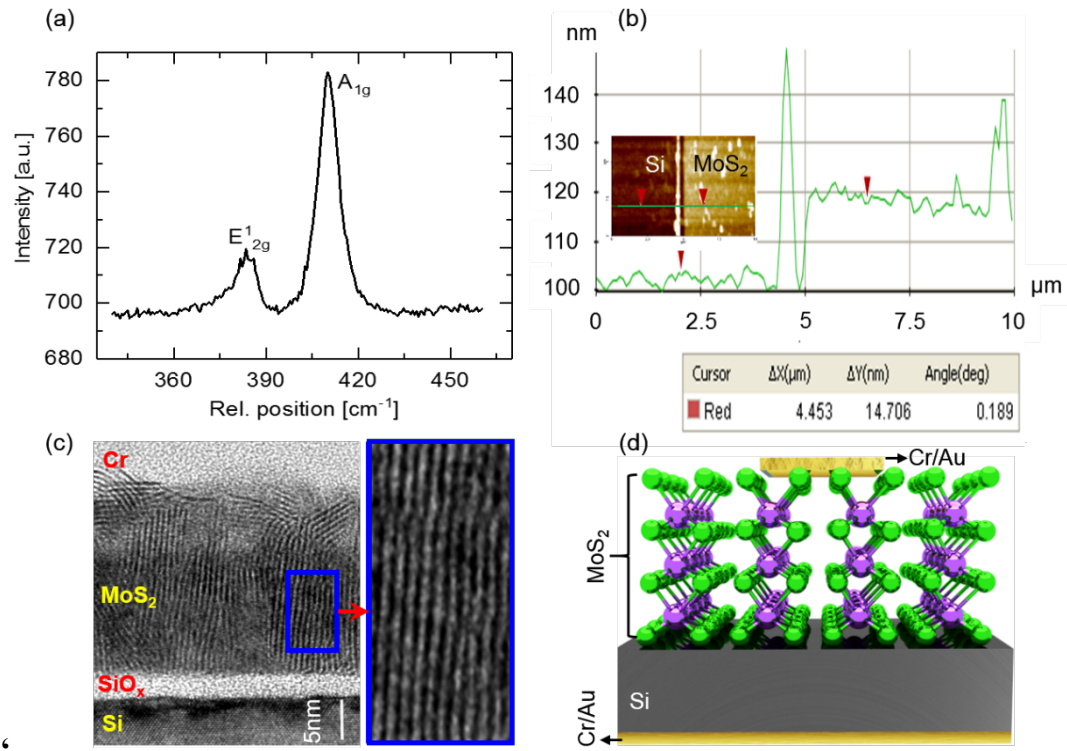

Figure 4- (a) Raman spectra with the $E^1_{2g}$ and $A_{1g}$ peaks at around 384 cm$^{-1}$ and 410 cm$^{-1}$ confirming the 2H-MoS$_2$ phase. The large $A_{1g}/E^1_{2g}$ intensity ratio indicates the formation of vertically aligned layers. (b) AFM scan of the MoS$_2$ film thickness. The inset shows the corresponding topography image from an AFM area scan distinctly delineating the MoS$_2$ and Si regions. The green line indicates the line profile along which the height difference was measured. (c) Cross-sectional TEM of the SSM structure revealing the nanocrystalline MoS$_2$ with vertical orientation of layers and the SiO$_x$ interfacial layer. The TEM image to the right shows a magnified view of the MoS$_2$ layers in the region indicated by the blue box in (c). (d) Schematic diagram illustrating the structure of the present memristive device with vertically aligned MoS$_2$ layers.